\documentclass{sf2a-conf2023}
\usepackage{graphicx}
\usepackage{hyperref}
\usepackage[]{natbib}  
\usepackage{epstopdf}

\def\BibTeX{{\rm B\kern-.05em{\sc i\kern-.025em b}\kern-.08em
    T\kern-.1667em\lower.7ex\hbox{E}\kern-.125emX}}
\bibpunct{(}{)}{;}{a}{}{,}  

\begin{document}

\TitreGlobal{SF2A 2023}


\title{Follow-up of gravitational waves alerts with IACTs using Astro-COLIBRI}

\runningtitle{Follow-up of gravitational waves alerts with IACTs using Astro-COLIBRI}

\author{Mathieu de Bony de Lavergne}\address{IRFU, CEA, Université Paris-Saclay, F-91191 Gif-sur-Yvette, France}
\author{Halim Ashkar}\address{Laboratoire Leprince-Ringuet, Ecole Polytechnique, CNRS, Institut Polytechnique de Paris, Palaiseau,
France}
\author{Atilla Kaan Alkan$^1$}\address{Université Paris-Saclay, CNRS, Laboratoire intersdisciplonaire des sciences du numérique, 91045, Orsay France}
\author{Jayson Mourier$^1$}
\author{Patrick Reichherzer}\address{Department of Physics, University of Oxford, Oxford OX1 3PU, United Kingdom}
\author{Fabian Schüssler$^1$}
\author{Monica Seglar-Arroyo}\address{Institut de Física d’Altes Energies (IFAE), Barcelona Institute of Science and Technology, E-08193
Barcelona, Spain}




\setcounter{page}{237}


\maketitle


\begin{abstract}

Follow-up of gravitational wave alerts has proven to be challenging in the past due to the large uncertainty on the localisation, much larger than the field of view of most instruments. A smart pointing strategy helps to increase the chance of observing the true position of the underlying compact binary merger event and so to detect an electromagnetic counterpart.

To tackle this, a software called tilepy has been developed and was successfully used by the H.E.S.S. collaboration to search for very-high energy gamma-ray emission from GWs during the O2 and O3 runs. The optimised tiling strategies implemented in tilepy allowed H.E.S.S. to be the first ground facility to point toward the true position of GW 170817. Here we will present the main strategy used by the software to compute an optimal observation schedule.

The Astro-COLIBRI platform helps to plan follow-up of a large range of transient phenomena including GW alerts. The integration of tilepy in this tool allow for an easy planning and visualisation of of follow-up of gravitational wave alert helping the astronomer to maximise the chance of detecting a counterpart. The platform also provides an overview of the multi-wavelength context by grouping and visualising information coming from different observatories alongside GW alerts.

\end{abstract}



\section{Introduction}

Since the first detection of gravitational waves in 2015, the LIGO and Virgo interferometers have identified more than a hundred of mergers. Despite continuous enhancements to the detectors, the localisation of these events remain uncertain, with uncertainties spanning hundreds to even thousands of square degrees. This challenge is particularly pronounced when attempting to follow up on these gravitational wave alerts using Imaging Atmospheric Cherenkov Telescopes (IACTs), which typically have a limited field of view, typically covering only a few tens of square degrees.

To increase the likelihood of accurately observing the event's precise location, a smart pointing strategy is essential. This strategy involves observing multiple sky fields, often referred to as tiles. To facilitate the scheduling and computation of these tiles, a software tool named {\em tilepy} has been developed and seamlessly integrated into the Astro-COLIBRI platform. This platform serves as a comprehensive solution for collecting alerts from various channels and efficiently scheduling observations, especially in the context of gravitational wave events, where {\em tilepy} plays a pivotal role.
  
\section{{\em tilepy}: A software for efficient gravitational wave follow-up}

\subsection{Working principle}

{\em tilepy} was originally developed for the purpose of following up on Gravitational Wave (GW) events using the H.E.S.S. telescopes \citep{2021JCAP...03..045A}. Especially the tool allowed H.E.S.S. to be first ground facility pointing towards the real position of GW 170817 \citep{Abdalla_2017}, the first coincidence event between gravitational wave and electromagnetic signal. The software is also used by LST-1, the first telescope of CTA, for the follow-up of gravitational waves during the O4 observation run \citep{2022icrc.confE.838C}. Today, the code of {\em tilepy} is accessible to everyone through its GitHub repository at the following address: \url{https://github.com/astro-transients/tilepy}. Furthermore an API to allow an easy access is accessible at the following address : \url{https://tilepy.com}.

{\em tilepy} operates on the basis of a grid of pointing directions that cover the entire sky. Its primary objective is to optimize scheduling to maximize coverage within the uncertainty region of a GW event by choosing the best pointing possible. Two main strategies are implemented in {\em tilepy} to achieve this goal.

The first strategy, called "2D" approach, directly utilizes the sky localization map provided by LIGO/Virgo. {\em tilepy}'s algorithm searches for the pointing direction that provides the most extensive coverage of the uncertainty region. Once identified, this position is removed from the map, and a new pointing direction can be computed in a similar manner. It's worth noting that this method can also be adapted for other types of alerts, such as Gamma-Ray Bursts (GRBs) detected by Fermi/GBM.

Gravitational wave interferometers not only provide sky localizations but also estimate the distance at each position in the sky, enabling a so called "3D" algorithm. This distance information can be cross-referenced with a galaxy catalog, with the current catalog of choice being GLADE+ \citep{10.1093/mnras/stac1443}. {\em tilepy} assigns a probability to each galaxy and then searches for the field of view that encompasses the highest cumulative probabilities by summing the probabilities of all galaxies within it. It is important to note that this strategy is applicable only up to distances of a few hundred megaparsecs, as the completeness of galaxy catalogs diminishes for larger distances.

For a more in-depth understanding of the algorithms used in {\em tilepy}, please refer to the detailed description provided in \citet{2021JCAP...03..045A}.

\subsection{An HTTP API for an easy use of {\em tilepy}}

To simplify the use of {\em tilepy}, a REstful API has been developed. This API allows users to perform tiling for a specific alert through an easy to use and publicly available endpoint. Most of {\em tilepy}'s essential parameters, such as telescope location, size of the field of view, maximum zenith angle, moon separation, and more, can be configured via the parameters when making a query. The results are returned in JSON format to be easily parsed by an automatic system. The API is accessible with it's documentation at the following address: \url{https://tilepy.com}. The system is currently hosted on AWS, which assures a high availability.

\section{Astro-COLIBRI, a tool for transient astronomy}

\subsection{The tool for transient astronomy}

\begin{figure}[ht!]
\centering
\begin{minipage}{.55\textwidth}
  \centering
  \includegraphics[width=0.9\textwidth,clip]{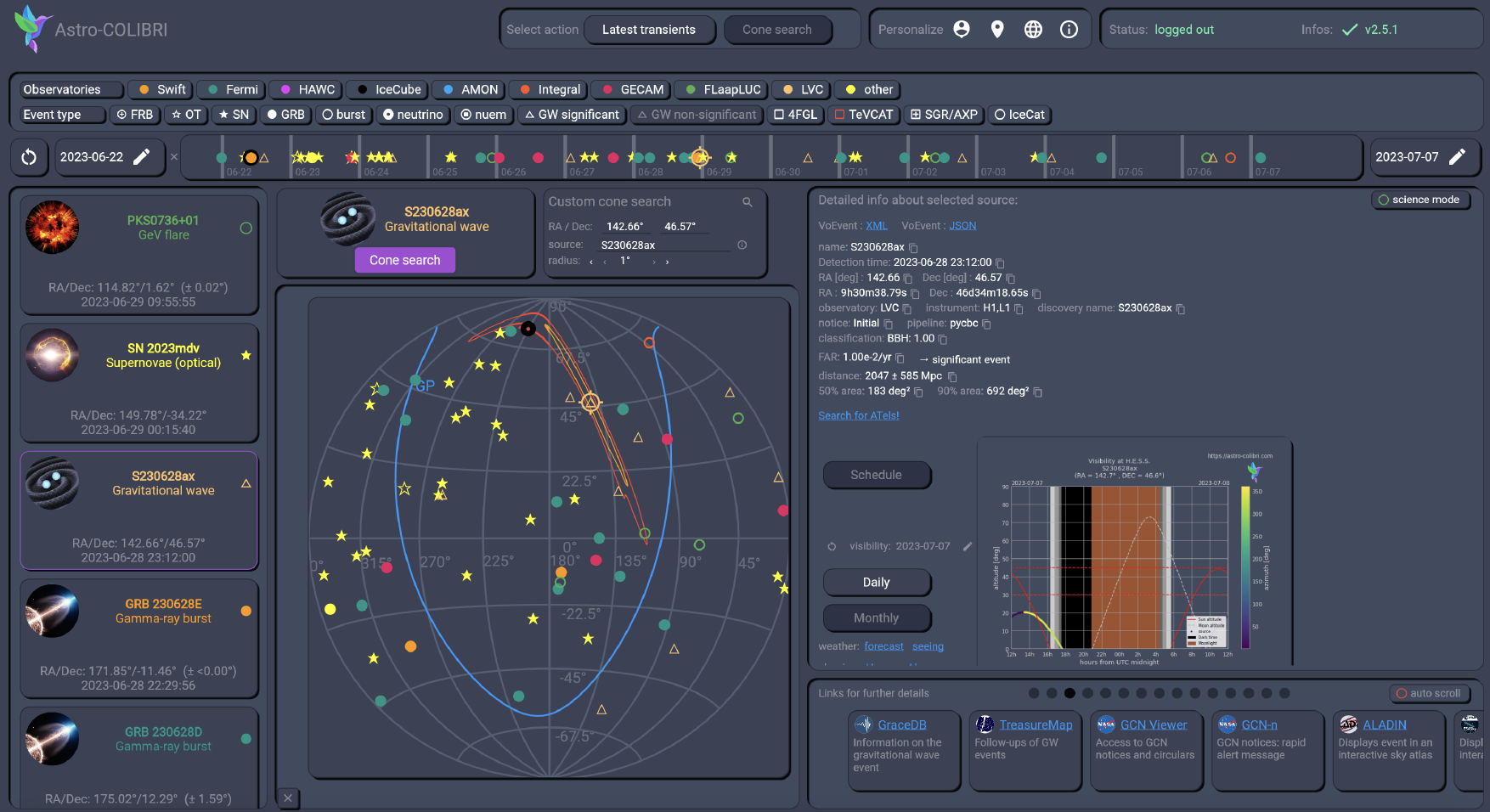}      
  \caption{Web interface of Astro-COLIBRI}
  \label{fig:astrocolibri_web}
\end{minipage}%
\begin{minipage}{.45\textwidth}
  \centering
  \includegraphics[width=0.8\textwidth,clip]{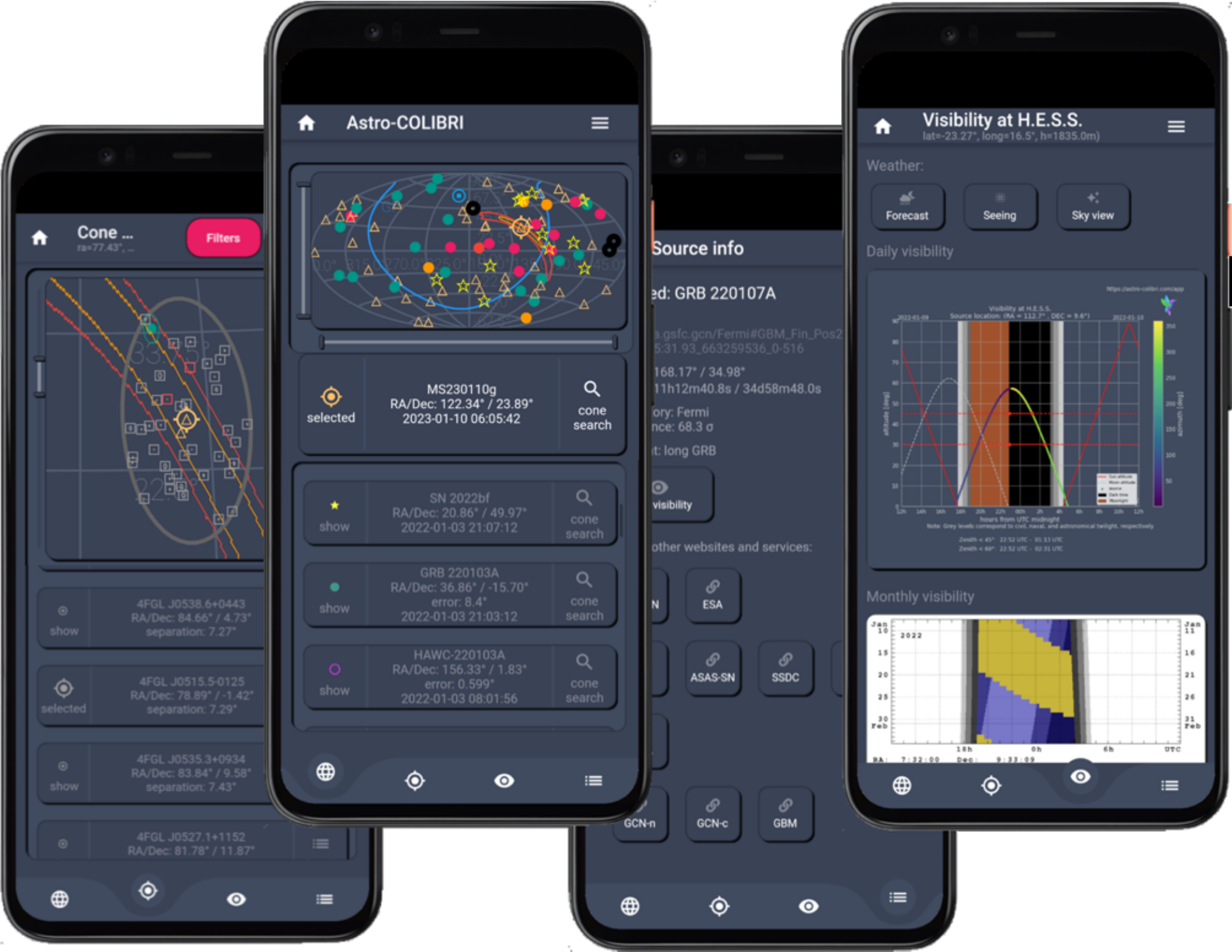}      
  \caption{Smartphone interface of Astro-Colibri}
  \label{fig:astrocolibri_app}
\end{minipage}
\end{figure}

Astro-COLIBRI is a platform for simplifying the study of transient astronomical events. It is integrating real-time multi-messenger observation tools in a simple and user-friendly graphical interface. This platform collects and consolidates information and alerts from various channels, providing astronomers with a comprehensive overview of the current transient phenomena across the sky and facilitating observation planning.

Astro-COLIBRI accommodates a wide range of astrophysical phenomena, making it a versatile tool for astronomers. It includes Active Galactic Nuclei (AGN), Gamma-ray Bursts (GRBs), Fast Radio Bursts (FRBs), Gravitational Waves (GWs), High-energy Neutrinos, Optical Transients (OT), Supernovae (SN), ....

The platform is accessible through two main interfaces. Firstly, a RESTful API hosted on AWS provides programmatic access to its features. You can access this API at the following address: \url{https://astro-colibri.science}, where comprehensive documentation is also available. Furthermore, Astro-COLIBRI offers a user-friendly graphical interface, which can be accessed through a Web browser (Figure \ref{fig:astrocolibri_web}) or a dedicated smartphone application (Figure \ref{fig:astrocolibri_app}). This last option also allows one to receive notification on your smartphone when a specific category of events is detected. 

A more detailed presentation of Astro-COLIBRI and together with the presentation of other use cases other than gravitation waves can be found in \citep{Schüssler:20235k}.

\subsection{Integration of {\em tilepy}}

\begin{figure}[ht!]
\centering
\begin{minipage}{.49\textwidth}
  \centering
  \includegraphics[width=0.99\textwidth,clip]{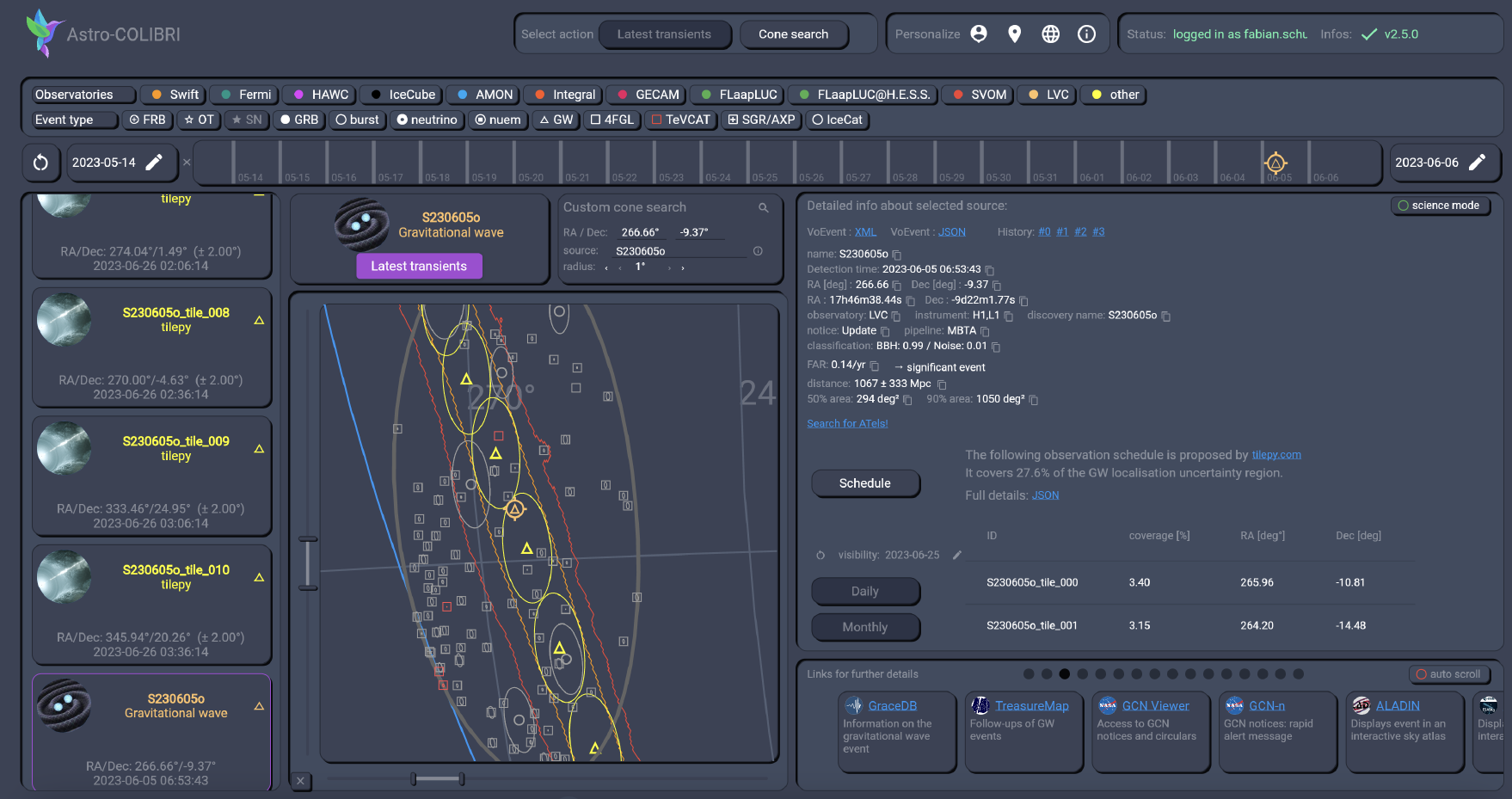}      
  \caption{Web interface of Astro-COLIBRI when using the tiling function. The maps on the left shows the position uncertainty contour of the gravitational wave alert alongside with the position of each tile. On the right, a table with the detailed information about each tile, including the time at which it should be observed is given.}
  \label{fig:astrocolibri_tilepy_web}
\end{minipage}
\vspace{0.05\linewidth}
\begin{minipage}{.45\textwidth}
  \centering
  \includegraphics[width=0.39\textwidth,clip]{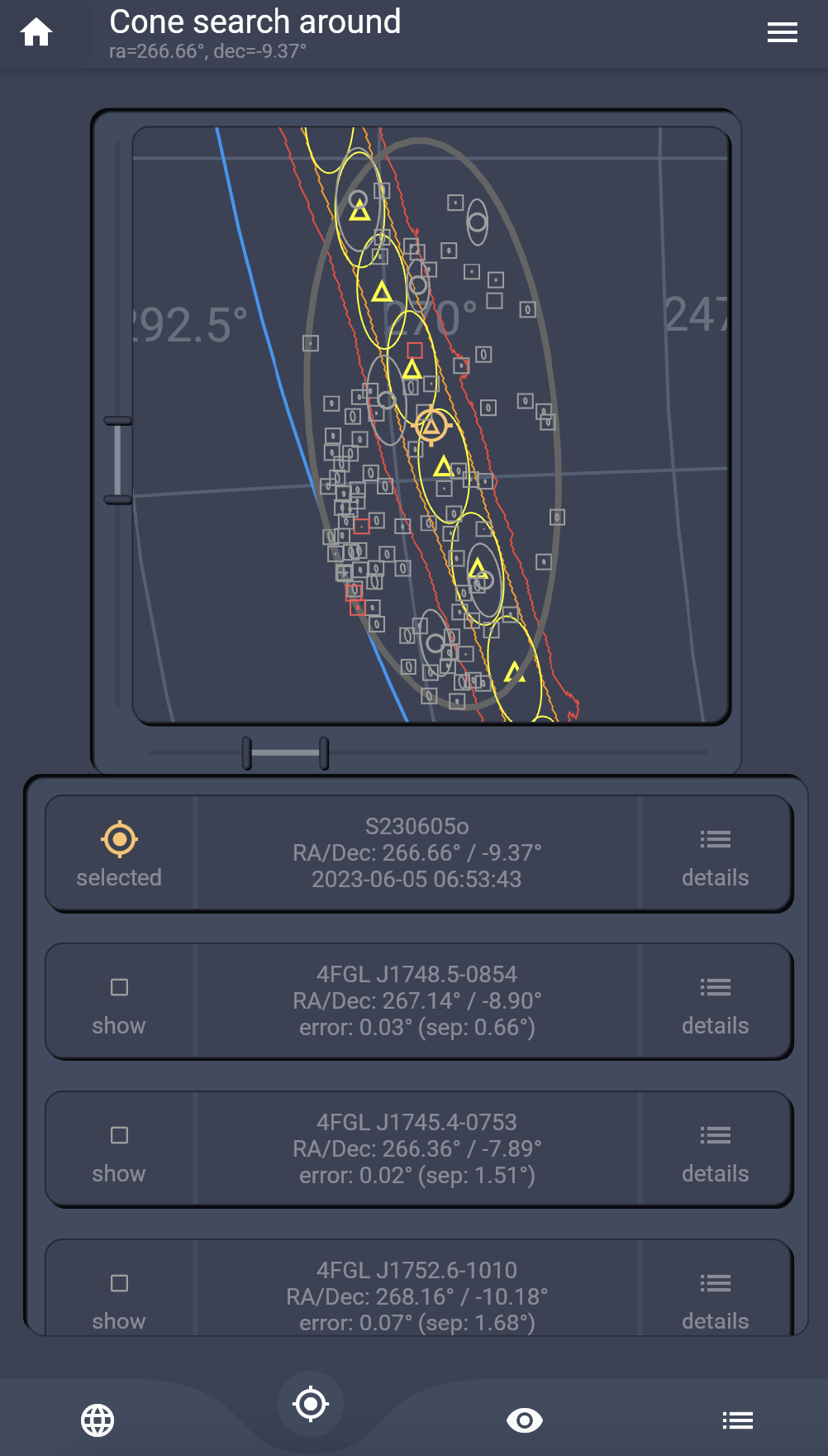}      
  \caption{Smartphone interface of Astro-COLIBRI when using the tiling function. At the top a map illustrates the position uncertainty contour of the gravitational wave alert alongside the position of each observation tile. The latter are also given in the list below.}
  \label{fig:astrocolibri_tilepy_app}
\end{minipage}
\end{figure}

To enhance the capabilities of Astro-COLIBRI to facilitate the follow-up of gravitational wave events, we have integrated the {\em tilepy} API into the Astro-COLIBRI platform. This integration simplifies the process of requesting and visualizing tile-based observations. When using this functionality, users can retrieve detailed information about each tile, including its optimal observation time. The platform also provides a visual representation of the uncertainty contours of the gravitational wave alongside the positions of each tile (cf. Figure \ref{fig:astrocolibri_tilepy_web}). Users of the smartphone interfaces can enjoy a similar experience through the available apps (Figure \ref{fig:astrocolibri_tilepy_app}).

\section{Conclusion}

In this contribution, we have presented Astro-COLIBRI, an advanced platform for real-time multi-messenger astrophysics. Astro-COLIBRI integrates multi-messenger observation tools into a comprehensive and user-friendly graphical interface, empowering astronomers to swiftly access critical information related to transient events.

One notable feature is the platform's ability to visualize recent gravitational wave events. Furthermore, with the recent integration of {\em tilepy}, astronomers can efficiently plan follow-up observations, significantly enhancing their capabilities in responding to these phenomena.

{\em tilepy} is a powerful tool for scheduling follow-up observations of GW events. It offers diverse strategies to optimize the coverage of the localization region, taking into account telescope specifications and visibility conditions. As an open source project, {\em tilepy} is accessible to a large number of researchers. We firmly believe that {\em tilepy} has the potential to become an indispensable resource for multi-messenger follow-up observations. With continued community-driven development, it could play a pivotal role in increasing the likelihood of detecting counterparts to GW events.

{\em tilepy} and Astro-COLIBRI are still under active development, with various planned enhancements. These include the incorporation of novel strategies to further refine observation scheduling within {\em tilepy} and improved filtering and notification capabilities for Astro-COLIBRI.

The Astro-COLIBRI and {\em tilepy} development team welcomes comments and feedback from the community to further improve the platform and can be contacted, respectively, at \href{mailto:astro.colibri@gmail.com}{astro.colibri@gmail.com} and \href{mailto:astro.tilepy@gmail.com}{astro.tilepy@gmail.com}.

\begin{acknowledgements}
The authors acknowledge the support of the French Agence Nationale de la Recherche (ANR) under reference ANR-22-CE31-0012. This work was also supported by the Programme National des Hautes Energies of CNRS/INSU with INP and IN2P3, co-funded by CEA and CNES and we acknowledge support by the European Union’s Horizon 2020 Programme under the AHEAD2020 project (grant agreement n. 871158). MSA acknowledges the support of Grant FJC2020-044895-I funded by MCIN/AEI/10.13039/501100011033 and by the European Union NextGenerationEU/PRTR.
\end{acknowledgements}

\bibliographystyle{aa}  
\bibliography{debony} 

\end{document}